\documentclass[12pt, oneside]{article}   	
\usepackage{geometry} 
\usepackage{amsmath} 								
\usepackage{graphicx}	
\usepackage{amssymb}
\usepackage{cite}
\usepackage{color,xspace}
\usepackage{extarrows}

\numberwithin{equation}{section}
\DeclareSymbolFont{extraup}{U}{zavm}{m}{n}
\DeclareMathSymbol{\vardiamond}{\mathalpha}{extraup}{87}
\usepackage{float}
\restylefloat{table}
\allowdisplaybreaks

\def\twomat[#1,#2][#3,#4]{\left( \begin{array}{cc} #1 & #2 \\ #3 & #4 \end{array} \right)}
\def\thv[#1,#2,#3]{\left( \begin{array}{c} #1 \\ #2 \\ #3 \end{array} \right)}
\def\twv[#1,#2]{\left( \begin{array}{c} #1 \\ #2 \end{array} \right)}
\def\lagrange{\mathcal{L}}
\def\nn{\nonumber}

\usepackage{color,xspace}
\usepackage{extarrows}
\def\MSUSY{\ensuremath{M_{\rm SUSY}}\xspace}

\title{Predicting Alignment in a Two Higgs Doublet~Model~$^{\dagger}$}

\date{}

\begin{document}

\begin{flushright}
\end{flushright}
\begin{center}

\vspace{1cm}
{\LARGE{\bf Predicting Alignment in a Two Higgs Doublet~Model}}\let\thefootnote\relax\footnote{Presented at the 7th International Conference on New Frontiers in Physics (ICNFP 2018), Crete, Greece, 4--12 July 2018.}

\vspace{1cm}

\large{\bf Karim Benakli$^\spadesuit$ \let\thefootnote\relax\footnote{$^\spadesuit$kbenakli@lpthe.jussieu.fr}
Yifan Chen$^{\vardiamond}$ \let\thefootnote\relax\footnote{$^\vardiamond$yifan.chen@lpthe.jussieu.fr}
 Ga\"etan~Lafforgue-Marmet$^\clubsuit$ \footnote{$^\clubsuit$glm@lpthe.jussieu.fr}
 \\[5mm]}

%{\small
%{\color{blue} \emph {1-- Sorbonne Universit\'es, UPMC Univ Paris 06, UMR 7589, LPTHE, F-75005, Paris, France \\
%2-- CNRS, UMR 7589, LPTHE, F-75005, Paris, France }}
{ \sl Laboratoire de Physique Th\'eorique et Hautes Energies (LPTHE),\\ UMR 7589,
Sorbonne Universit\'e et CNRS, 4 place Jussieu, 75252 Paris Cedex 05, France.}

\end{center}
%\vspace{0.7cm}
% Contact information of the corresponding author
%\corres{Correspondence: kbenakli@lpthe.jussieu.fr}

% Current address and/or shared authorship
%\firstnote{Current address: Affiliation 3} 
%\secondnote{These authors contributed equally to this work.}
% The commands \thirdnote{} till \eighthnote{} are available for further notes
%\footnote{Presented at the 7th International Conference on New Frontiers in Physics (ICNFP 2018),  Crete, Greece, 4--12 July 2018.}
%\simplesumm{} % Simple summary

%\conference{} % An extended version of a conference paper

% Abstract (Do not insert blank lines, i.e. \\) 
\abstract{We show that a non-abelian global $SU(2)_R$ R-symmetry acting on the quartic part of the two Higgs Doublet Model leads, at tree-level, to an automatic alignment without decoupling. An example of phenomenologically viable model with this feature is the the low energy effective field theory of the Minimal Dirac Gaugino Supersymmetric Model in the limit where the adjoint scalars are decoupled. We discuss here how the $SU(2)_R$ can be identified with the R-symmetry of the $N=2$ supersymmetry in the gauge and Higgs sectors. We also review how the radiative corrections lead to a very small misalignment.}

% Keywords
%\keyword{extended supersymmetry,  dirac gauginos,  extended higgs sectors}

%%%%%%%%%%%%%%%%%%%%%%%%%%%%%%%%%%%%%%%%%%
%\begin{document}

\section{Introduction}

The Standard Model Higgs is the only known fundamental spin zero particle in Nature. The existence of additional fundamental scalars is not excluded and happens in Early Universe cosmological and supersymmetric models. Such additional  scalars could mix with the observable Higgs. This leads to strong constraints from existing experimental data. In particular, this requires that  the observed Higgs is \emph {aligned} with the direction acquiring a non-zero vacuum expectation value (v.e.v).  This can be achieved by decoupling the additional scalars by making them heavy enough. However, the alignment can also be  a consequence of a symmetry of the model in with case the new scalar masses could lie in a range within the reach of future searches at the LHC . Such an   \emph {alignment without decoupling}~\cite{Gunion:2002zf} was realized in \cite{Antoniadis:2006uj} (
see the discussion of the spectrum in Section \ref{sec3} of \cite{Antoniadis:2006uj},  the model was not engineered for this purpose but as a scenario for supersymmetry breaking, and therefore we can consider the alignment there as a "prediction'' of the model) and  discussed  later in \cite{Ellis:2016gxa,Benakli:2018vqz}. In particular, it was shown in  \cite{Benakli:2018vqz} that the alignment survives with an impressive precision when radiative corrections are taken into account. The mechanisms behind this successful alignment are a combination of a global $SU(2)_R$ symmetry of the quartic potential and diverse cancellations due to supersymmetry as discussed in \cite{Benakli:2018vjk} and will be reviewed here.

The  scalar potential of \cite{Antoniadis:2006uj}, studied  in \cite{Belanger:2009wf}, is that of a Two Higgs Doublet Model (2HDM)   (
for an introduction to 2HDM, see for example \cite{Gunion:1989we,Branco:2011iw,Djouadi:2005gj}). Alignment is not necessarily due to symmetries. Viable cases have been discussed for example in \cite{Bernon:2015qea,Bernon:2015wef,Carena:2013ooa,Carena:2015moc,Haber:2017erd} for the MSSM 
and NMSSM.  However this looks as an ad-hoc specific choice of the model parameters.  One could search for symmetries of the 2HDM (e.g.,  \cite{Davidson:2005cw,Ivanov:2005hg,Ferreira:2009wh}) that imply alignment without decoupling \cite{Dev:2014yca,Lane:2018ycs}. Quite often they lead to problematic phenomenological consequences, as massless quarks \cite{Ferreira:2010bm}. In the supersymmetric model of \cite{Antoniadis:2006uj}, the alignment at tree-level is also a prediction of a symmetry: a non abelian R-symmetry. However, this symmetry acts only on part of the Lagrangian and does not lead to phenomenological issues.    %

In \cite{Antoniadis:2006uj}, the (non-chiral) gauge and Higgs states appear in an $N=2$ supersymmetry sector while the matter states, quarks and leptons, appear in an $N=1$ sector. Early models suffered from the non-chiral nature of quarks and leptons \cite{Fayet:1975yi,delAguila:1984qs} as they have required that $N=2$ supersymmetry acts on the whole SM states.    An important feature of \cite{Antoniadis:2005em,Antoniadis:2006eb,Antoniadis:2006uj,Allanach:2006fy,Belanger:2009wf} is that  gauginos have Dirac masses~\cite{Fayet:1978qc,Polchinski:1982an,Hall:1990hq,Fox:2002bu,Benakli:2008pg}. The $N=2$ extension have implication for Higgs boson physics as discussed in \cite{Belanger:2009wf,Benakli:2009mk,Amigo:2008rc,Benakli:2010gi,Choi:2010gc,Benakli:2011vb,Benakli:2011kz,Itoyama:2011zi,Benakli:2012cy,Benakli:2014cia,Martin:2015eca,Braathen:2016mmb,Unwin:2012fj,Chakraborty:2018izc,Csaki:2013fla,Benakli:2016ybe,Nelson:2015cea,Alves:2015kia,Alves:2015bba}. We will review here how this alignment emerges and  how higher order corrections induce a small misalignment.  
 
%%%%%%%%%%%%%%%%%%%%%%%%%%%%%%%%%%%%%%%%%%
%--------------------------------------------------------------
\section{Higgs Alignment from an \boldmath{$SU(2)$} Symmetry}
%\label{2HDM_Limit}
%---------------------------------------------------------

We review here how to obtain alignment as a consequence of an $SU(2)$ symmetry acting on the quartic potential. Alignment as consequence of these relations between the different dimensionless coupling is trivial and has been discussed by many authors, for example in \cite{Gunion:2002zf,Dev:2014yca}. However, these works looked at symmetries of the whole Lagrangian and therefore they explicitly associate the obtained alignment with equalities or vanishing squared-mass parameters. Here, the situation is a bit different. By construction as we will discuss below, our $SU(2)$ symmetry does not, and \textit{can not}, act on the quadratic part of the potential in contrast with previous works assumptions. We need to explain why, still, this is enough to imply alignment. However, most useful to us, we want to read  the amount of misalignment as function of the decomposition under $SU(2)$ of the quartic potential \cite{Benakli:2018vjk}.

The standard parametrization of a generic 2HDM is  (
Here, for simplicity, we assume CP conservation. All couplings and vacuum expectation values are assumed to be real.): 
\begin{eqnarray}
V_{EW} &=& V_{2\Phi} +V_{4\Phi} 
\label{decomp2HDM} 
\end{eqnarray}
where
\begin{eqnarray}
V_{2\Phi} &=& m_{11}^2 \Phi_1^\dagger \Phi_1 + m_{22}^2 \Phi_2^\dagger \Phi_2 - [m_{12}^2 \Phi_1^\dagger \Phi_2 + \text{h.c}] \nonumber \\
V_{4\Phi} &=& \frac{1}{2} \lambda_1 (\Phi_1^\dagger \Phi_1)^2 + \frac{1}{2} \lambda_2 (\Phi_2^\dagger \Phi_2)^2 \nonumber \\
& & +  \lambda_3(\Phi_1^\dagger \Phi_1) (\Phi_2^\dagger \Phi_2) + \lambda_4 (\Phi_1^\dagger \Phi_2)(\Phi_2^\dagger \Phi_1) \nonumber \\ 
& & + \left[ \frac{1}{2} \lambda_5 (\Phi_1^\dagger \Phi_2)^2 + [\lambda_6 (\Phi_1^\dagger \Phi_1) + \lambda_7 (\Phi_2^\dagger \Phi_2)] \Phi_1^\dagger \Phi_2 + \text{h.c} \right]\,,  
\label{reparametriz2HDM} 
\end{eqnarray}
We expect the parameters $\lambda_i$ to contain leading order tree-level values  with  corrections from loops $\delta \lambda_i^{(rad)}$ but also at tree-level $\delta \lambda_i^{(tree)}$ from  threshold corrections  due to integration of heavy states:
\begin{align}
\lambda_i =&\lambda_i^{(0)} + \delta \lambda_i^{(tree)} + \delta \lambda_i^{(rad)}
\label{deltaLs}
\end{align}

Now, put the two Higgs doublets together in a bi-doublet  $(\Phi_1, \Phi_2)^T$  where $\Phi_1$ and $\Phi_2$ can be represented as columns with two entries. We then consider the $SU(2)$ symmetry that rotates the two doublets among themselves, therefore acting horizontally. We denote this group as $SU(2)_R$ (R stands for R-symmetry as we will see below) and the two fields appear now in the fundamental representation of the $SU(2)_R$. 

A potential that is invariant under $SU(2)_R$ will contain only singlets of $SU(2)_R$ and can be written~as:
\begin{eqnarray}
V_{4\Phi} &=& \, \, \lambda_{|0_1,0>}   |0_1,0\rangle  \quad +  \quad  \lambda_{|0_2,0>}   |0_2,0\rangle
\label{reparam2HDM-SU2R} 
\end{eqnarray}
where $|l,m>$ are the spin representation of $SU(2)_R$ in the standard notation. It is easy to check that:
\begin{eqnarray}
\begin{array}{lll}
|0_1,0\rangle&=& \frac{1}{2}\left[(\Phi^\dagger_1\Phi_1) + (\Phi^\dagger_2 \Phi_2)\right]^2 \, ,
\end{array}
\label{1-1ofSU(2)}
\end{eqnarray}
and 
\begin{eqnarray}
\begin{array}{lll}
|0_2,0\rangle&=& - \frac{1}{\sqrt{12}}\left[\left((\Phi^\dagger_1\Phi_1) - (\Phi^\dagger_2 \Phi_2)\right)^2 \, 
+ 4(\Phi^\dagger_2\Phi_1)(\Phi^\dagger_1\Phi_2)\right]
\end{array}
\label{1-2ofSU(2)}
\end{eqnarray}
while comparing with (\ref{reparametriz2HDM}) gives:
\begin{eqnarray}
\lambda_{|0_1,0>} &=&\frac {\lambda_1 + \lambda_2 + 2\lambda_3}{ 4} 
\label{lambdaofSU(2)1}
\end{eqnarray}
and

\begin{eqnarray}
\lambda_{|0_2,0>} = -\frac  {\lambda_1 + \lambda_2 - 2\lambda_3 +4\lambda_4 }{ 4\sqrt{3}} 
\label{lambdaofSU(2)2}
\end{eqnarray}
The absence of other $|l,m>$'s  can be enforced by choosing
\begin{align}
\lambda_5 =& \lambda_6 = \lambda_7 =0.
\label{lambdaZeros}
\end{align}
For the case of CP conserving Lagrangian under consideration, there are two CP even scalars with squared-mass matrix in the Higgs basis (e.g., \cite{Davidson:2005cw})
\begin{eqnarray}
\mathcal{M}^2_h = \begin{pmatrix}
Z_1 v^2 & Z_6 v^2 \\
Z_6 v^2 & m_A^2 + Z_5 v^2 \end{pmatrix} \, . 
\label{2HDM_mass_matrix}
\end{eqnarray}
These are given by
\begin{align}
Z_1 =& \lambda_1c_\beta^4 + \lambda_2 s_\beta^4 + \frac{1}{2} \lambda_{345} s_{2\beta}^2,   \nn\\
 Z_5 =& \frac{1}{4} s_{2\beta}^2 \left[ \lambda_1 + \lambda_2 - 2\lambda_{345}\right] + \lambda_5  \nn\\
Z_6 =& -\frac{1}{2} s_{2\beta} \left[\lambda_1 c_\beta^2 - \lambda_2 s_\beta^2 - \lambda_{345} c_{2\beta} \right] 
\label{2HDMZ}
\end{align} 
where $\lambda_{345} \equiv \lambda_3 + \lambda_4 + \lambda_5$, while  the pseudo-scalar mass $m_A$ is given by
\begin{align}
m_A^2 =& - \frac{m_{12}^2}{s_\beta c_\beta} - \lambda_5 v^2 \qquad \xlongrightarrow{ \lambda_5=0} \qquad  - \frac{m_{12}^2}{s_\beta c_\beta}
\end{align}
Here, we have defined :
\begin{eqnarray}
<Re(\Phi_{2}^0)>& =& {v s_\beta }, \qquad <Re(\Phi_{1}^0)> ~=~{v c_\beta},  
\end{eqnarray}
where:
\begin{eqnarray}
c_\beta &\equiv& \cos \beta,\qquad   s_\beta ~\equiv~ \sin \beta, \qquad  t_\beta ~\equiv ~\tan\beta \, ,  \qquad   0 \leqslant \beta \leqslant \frac{\pi}{2} \nonumber \\
c_{2\beta} &\equiv& \cos 2\beta,\qquad   s_{2\beta}~ \equiv~\sin 2\beta
\end{eqnarray}

The off-diagonal squared-mass matrix element $Z_6$ measures the displacement from alignment. It~can be written in the $SU(2)_R$  basis as
\begin{align}
Z_6 = \frac{1}{2} s_{2\beta} \left[  \sqrt{2} \lambda_{|1,0>} -  \sqrt{6} \lambda_{|2,0>} c_{2\beta}  + (\lambda_{|2,-2>} + \lambda_{|2,+2>}) c_{2\beta}. \right] 
\label{Z6-tree-SU(2)}
\end{align} 
where we used the notation (see \cite{Ivanov:2005hg}):
\begin{eqnarray}
\begin{array}{lll}
|1,0\rangle&=& \frac{1}{\sqrt{2}}\left[(\Phi^\dagger_2\Phi_2) - (\Phi^\dagger_1 \Phi_1)\right]
\left[(\Phi^\dagger_1\Phi_1) + (\Phi^\dagger_2 \Phi_2)\right] \\[1.5mm]
|2,0\rangle&=& \frac{1}{\sqrt{6}}\left[(\Phi^\dagger_1\Phi_1)^2 + (\Phi^\dagger_2 \Phi_2)^2 
 - 2(\Phi^\dagger_1\Phi_1)(\Phi^\dagger_2\Phi_2) -  2(\Phi^\dagger_1\Phi_2)(\Phi^\dagger_2\Phi_1)\right] \\[1.5mm]
|2,+2\rangle&=& (\Phi^\dagger_2\Phi_1)(\Phi^\dagger_2 \Phi_1) \\ [1.5mm]
|2,-2\rangle&=& (\Phi^\dagger_1\Phi_2)(\Phi^\dagger_1 \Phi_2) 
\end{array}
\label{5ofSU(2)}
\end{eqnarray}
The coefficients appearing in (\ref{Z6-tree-SU(2)}) are given by:
\begin{eqnarray}
\lambda_{|1,0>} &=&  \frac {\lambda_2 - \lambda_1 }{ 2\sqrt{2}}\, \qquad \xlongrightarrow{SU(2)_R}  \qquad  0 \nn \\ [1.5mm]
\lambda_{|2,0>}  &= &\frac {\lambda_1 + \lambda_2 - 2\lambda_3 - 2\lambda_4 }{ \sqrt{24}}\, \qquad \xlongrightarrow{SU(2)_R}  \qquad  0  \nn  \\ [1.5mm]
\lambda_{|2,+2>} &=& \frac{\lambda_5^*}{2}\, \quad \xlongrightarrow{{\rm leading\,  order}}  \quad  0, \qquad 
 \lambda_{|2,-2>} = \frac{\lambda_5}{2}\, \quad \xlongrightarrow{{\rm leading\,  order}}  \quad  0.
\label{lambdaofSU(2)}
\end{eqnarray}
We see that the invariance under $SU(2)_R$ implies alignment.  The breaking of $SU(2)_R$ even just to its abelian sub-group spoils the alignment. Also, note that we have $\lambda_5 =0$, there is no contribution from $|2,\pm 2>$.

The quadratic part of the scalar potential can be written as:
\begin{eqnarray}
V_{2\Phi} &=& \, \,   \frac{ m_{11}^2  + m_{22}^2}{\sqrt{2}}\times  \frac{1}{\sqrt{2}} \left[  (\Phi_1^\dagger \Phi_1) + (\Phi_2^\dagger \Phi_2)  \right]
 \nonumber \\
& & +  \frac{ m_{11}^2  - m_{22}^2}{\sqrt{2}}\times  \frac{1}{\sqrt{2}} \left[  (\Phi_1^\dagger \Phi_1) - (\Phi_2^\dagger \Phi_2)  \right] \nonumber \\
& & - [m_{12}^2 \Phi_1^\dagger \Phi_2 + \text{h.c}] 
\label{reparam2HDM} 
\end{eqnarray}
where the only $SU(2)_R$ invariant part is given by the first line. The minimization of the potential leads to (e.g., \cite{Haber:1993an}):
\begin{eqnarray}
0 &=& m_{11}^2 - t_{\beta} m_{12}^2  + \frac{1}{2} v^2 c_\beta^2 (\lambda_1 +  \lambda_6 t_\beta + \lambda_{345} t_\beta^2 + \lambda_7 t_\beta^2)    \nn  \\
0 &=& m_{22}^2 - \frac{1}{t_{\beta}} m_{12}^2  + \frac{1}{2} v^2 s_\beta^2 (\lambda_2 +  \lambda_7 \frac{1}{t_\beta} + \lambda_{345} \frac{1}{t_\beta^2} + \lambda_6 \frac{1}{t_\beta^2})  
\label{EqtsMotion12} 
\end{eqnarray} 
Using that (\ref{lambdaofSU(2)}) implies $\lambda_1= \lambda_2 = \lambda_{345}\equiv \lambda$ and $\lambda_6= \lambda_7 = 0$, the equations (\ref{EqtsMotion12}) become: 
\begin{eqnarray}
0 &=& m_{11}^2 - t_{\beta} m_{12}^2  + \frac{1}{2} \lambda v^2     \\
0 &=& m_{22}^2 - \frac{1}{t_{\beta}} m_{12}^2  + \frac{1}{2} \lambda v^2  
\label{EqtsMotion2} 
\end{eqnarray} 
which subtracted one of the other give (for $s_{2\beta} \neq 0$) 
\begin{eqnarray}
0 &=& \frac{1}{2} (m_{11}^2 - m_{22}^2) s_{2\beta} + m_{12}^2  c_{2\beta} \equiv Z_6 v^2
\label{EqtsMotion3} 
\end{eqnarray} 
Thus the constraint of $SU(2)_R$ invariance of the quartic part of the potential implies an automatic alignment without decoupling.

%%%%%%%%%%%%%%%%%%%%%%%

\section{A Model with \boldmath{$SU(2)_R$} Symmetry}\label{sec3}

In the context of supersymmetric theories, one way to obtain the $SU(2)_R$ described above is to make of the two Higgs doublets one hypermultiplet $(\Phi_1, \Phi_2)^T$, the $SU(2)_R$ becomes an R-symmetry and supersymmetry is extended to $N=2$ in the Higgs sector. Now the $SU(2)_R$ R-symmetry  will act here as an $SU(2)$ Higgs family symmetry \cite{Davidson:2005cw,Ivanov:2005hg}, but now only on the quartic potential contains only terms that are invariant (singlet) under $SU(2)_R$.
As the Higgs doublets quartic potential receives contributions from  $D$-terms, we must also extend the $N=2$ supersymmetry to the gauge sector.  This implies the presence of chiral superfields in the adjoint representations of SM gauge group. These are a singlet $\mathbf{S}$ and an $SU(2)$ triplet  $\mathbf{T}$. We define
\begin{eqnarray}
S &=& \frac{S_R + iS_I}{\sqrt{2}} \\
T &=& \frac{1}{2} 
\begin{pmatrix}
T_0 & \sqrt{2} T_+ \\
\sqrt{2}T_- & -T_0 
\end{pmatrix} \,, \qquad T_i = \frac{1}{\sqrt{2}}\, (T_{iR} + i T_{iI})  \quad {\rm  with}  \quad i=0,+, -
\end{eqnarray}
They contribute to the superpotential  by promoting the gauginos to Dirac fermions, but also by generating new Higgs interactions through:
\begin{eqnarray}
W = && \sqrt{2} \, \mathbf{m}_{1D}^\alpha \mathbf{W}_{1\alpha} \mathbf{S} + 2 \sqrt{2} \, \mathbf{m}_{2D}^\alpha \text{tr} \left( \mathbf{W}_{2\alpha} \mathbf{T}\right)  + \frac{M_S}{2} \mathbf{S}^2+ \frac{\kappa}{2} \mathbf{S}^3 + M_T \, \text{tr} (\mathbf{TT}) \,  \nonumber \\ 
 && + \mu \,  \mathbf{H_u} \cdot \mathbf{H_d}+ \lambda_S S \, \mathbf{H_u} \cdot \mathbf{H_d} + 2 \lambda_T \, \mathbf{H_d} \cdot \mathbf{T H_u} \,,
\end{eqnarray} 
\noindent where the Dirac masses are parametrized by spurion superfields $\mathbf{m}_{\alpha i D} = \theta_\alpha m_{iD} $ where $\theta_\alpha$ are the Grassmannian superspace coordinates. The $\lambda_{S,T}$ are not arbitrary as $N=2$ supersymmetry  implies 
\begin{align}
\lambda_S= \frac{1}{\sqrt{2}} g_Y, \qquad \lambda_T =\frac{1}{\sqrt{2}} g_2
\label{LSTN2}
\end{align}
where $g_Y$ and $g_2$ stand for the hyper-charge and $SU(2)$ gauge couplings, respectively. The Higgs potential gets also contributions from soft supersymmetry breaking terms. We chose for simplicity the parameters to be real and we write
\begin{align}
\lagrange_{\rm soft} =& m_{H_u}^2 |H_u|^2 + m_{H_d}^2 |H_d|^2 + B{\mu} (H_u \cdot H_d + \text{h.c})  \nonumber \\ 
 & + m_S^2 |S|^2 + 2 m_T^2 \text{tr} (T^{\dagger} T) + \frac{1}{2} B_S \left(S^2 + h.c\right)+  B_T\left(\text{tr}(T T) + h.c.\right)  \label{soft} \\
& + A_S \left(S H_u \cdot H_d + h.c \right) + 2 A_T  \left( H_d \cdot T H_u + h.c \right) + \frac{A_\kappa}{3} \left( S^3 + h.c. \right) + A_{ST} \left(S \mathrm{tr} (TT) + h.c \right). \nn 
\end{align}

A peculiar 2HDM, with an extended set of light charginos and neutralinos, is obtained by integrating out of the adjoint scalars. The details of this potential were discussed in \cite{Belanger:2009wf}. The result can be mapped to (\ref{reparametriz2HDM}) after the identification
\begin{align}
\Phi_2 = H_u, \qquad \Phi_1^i = -\epsilon_{ij} (H_d^j)^* \Leftrightarrow \twv[H_d^0,H_d^-] = \twv[\Phi_1^0,-(\Phi_1^+)^*] 
\end{align}
from which we can now read
\begin{eqnarray}
m_{11}^2 &=& m_{H_{d}}^2 + \mu^2, \qquad m_{22}^2 ~= ~m_{H_{u}}^2 + \mu^2, \qquad m_{12}^2~ =~ B\mu . 
\label{2HDM_params}
\end{eqnarray}
and 
\begin{align}
\lambda_1^{(0)}= \lambda_2^{(0)} =& \frac{1}{4} (g_2^2 + g_Y^2)  \nn \\
\lambda_3^{(0)} =&   \frac{1}{4}(g_2^2 - g_Y^2) + 2 \lambda_T^2  \qquad \xlongrightarrow{N=2}  \qquad  \frac{1}{4}(5 g_2^2 - 
g_Y^2)  \nn  \\
\lambda_4 ^{(0)}=& -\frac{1}{2}g_2^2 + \lambda_S^2 - \lambda_T^2  \qquad \xlongrightarrow{N=2} \qquad - g_2^2 + \frac{1}{2} g_Y^2\nn\\
\lambda_5 =& \, \, \lambda_6 = \lambda_7 =0.
\label{EQ:MDGSSMTree}
\end{align}
as given in \cite{Belanger:2009wf,Benakli:2018vqz}.

Again, restricting to the case of CP conserving Lagrangian, the two CP even scalars have squared-mass matrix (\ref{2HDM_mass_matrix})
with
\begin{align}
Z_1 & \qquad \xlongrightarrow{N=2} \qquad \frac{1}{4} (g_2^2 + g_Y^2)  \nn\\
 Z_5 & \qquad \xlongrightarrow{N=2}  \qquad 0 \nn\\
Z_6 &  \qquad \xlongrightarrow{N=2} \qquad  0.
\label{2HDMZ}
\end{align} 
We use:
\begin{eqnarray}
M_Z^2 & =& \frac{g_Y^2 + g_2^2}{4} v^2 \, ,  \qquad  v \simeq 246 {\rm GeV}  \\
<H_{uR}>& =& {v s_\beta }, \qquad <H_{dR}>={v c_\beta},  \\
 <S_R>&=& v_s \, , \qquad <T_R>=v_t
\end{eqnarray}

Now $m_A$ is given by
\begin{align}
m_A^2 =& - \frac{m_{12}^2}{s_\beta c_\beta} - \lambda_5 v^2 \qquad \xlongrightarrow{N=2} \qquad  - \frac{m_{12}^2}{s_\beta c_\beta}
\end{align}
and squared-mass matrix has eigenvalues:
\begin{align}
m_{h}^2 &= \frac{1}{4} (g_2^2 + g_Y^2) v^2= M_Z^2 \nn\\
 m_{H}^2 &= m_A^2 
\label{ZH}
\end{align} 
while the charged Higgs has a mass
\begin{align}
m_{H^+}^2 =& \frac{1}{2} ( \lambda_5 - \lambda_4) v^2 + m_{A}^2 \qquad \xlongrightarrow{N=2} \qquad    \frac{1}{2}  (g_2^2 - \frac{1}{2} g_Y^2) v^2 + m_{A}^2=3M_W^2-M_Z^2+m_A^2.
\end{align}

Also, the leading-order squared-masses for the real part of the adjoint fields are \cite{Benakli:2011kz}:
\begin{align}
m_{SR}^2 =& m_S^2 + 4 m_{DY}^2 + B_S , \qquad m_{TR}^2 = m_T^2  + 4 m_{D2}^2 + B_T \, .
\label{STscalarmasses}
\end{align}
where we have taken $M_S=M_T=0$.

Let us turn now to the quadratic part of the potential. It can be written as (\ref{reparam2HDM}). Imposing a Higgs family symmetry would have required that both coefficients of the two $SU(2)_R$ non-singlets operators to vanish, therefore $m_{11}^2  = m_{22}^2$ and $m_{12}=0$. First, this would imply $m_{A}^2=0$ which is not a viable feature. Second, 
 the mass parameters in the quadratic potential under $SU(2)_R$ are controlled by the supersymmetry breaking mechanism and this is not expected to preserve the $R$-symmetry.  It was shown in \cite{Benakli:2008pg} that absence of tachyonic directions in the adjoint fields scalar potential implies that in a gauge mediation scenario that either breaking or messenger sectors should not be $N=2$ invariant. Thus, the quadratic potential can not be invariant under $SU(2)_R$.

%--------------------------------------------------------------
\section{\boldmath{$R$}-Symmetry Breaking and Misalignment}
%\label{2HDM_Limit}
%---------------------------------------------------------

%%%%%%%%%%%%%%%%%%%%%%%%

We have found above that invariance under $SU(2)_R$ symmetry of the quartic scalar potential is sufficient  to insure the Higgs alignment. This is because the symmetry relates different dimensionless couplings and forces others to vanish in such a way that $Z_6$ itself vanishes. However, this symmetry will be broken at least by quantum corrections to the mentioned set of couplings from sectors of the theory that do not respect the $SU(2)_R$ symmetry. Unexpectedly, it was found in  \cite{Benakli:2018vqz} that these corrections are very small. This was checked numerically including all threshold and two-loop effects when they are known. Here, we would like to exhibit the structure of these corrections with respect to group theoretical organization of the scalar potential in representations of $SU(2)_R$ .

We start by writing the quartic scalar potential as:
\begin{eqnarray}
V_{4\Phi} &=& \, \, \sum_{j,m} \lambda_{|j,m>} \times   |j,m\rangle
\label{reparam2HDM-2} 
\end{eqnarray}
where $|j,m\rangle$ are the irreducible representations of $SU(2)_R$.  

Here $\lambda_5 =0$, thus the misalignment is parametrized by
\begin{align}
Z_6 = \frac{1}{2} s_{2\beta} \left[  \sqrt{2} \lambda_{|1,0>} - \sqrt{6} \lambda_{|2,0>} c_{2\beta}  \right] 
\label{Z6-tree-SU(2)-2}
\end{align} 
We see that the conservation of the $U(1)_{R}^{(diag)}$ subgroup of $SU(2)_R$ is  \emph {not sufficient} for alignment as the presence of either of $|1,0\rangle$ or $|2,0\rangle$ leads to misalignment.

In our model $\lambda_{|i,0>}$ are corrections generated by higher order corrections to the tree-level $\lambda_{|i,0>}^{(0)}$.
First, there are tree-level corrections corresponding to thresholds when integrating out adjoint scalars. Note that the Higgs $\mu$-term and the Dirac masses $m_{1D}, m_{2D}$ are kept small , in the sub-TeV region. We~have:
\begin{align}
\delta \lambda_1^{(tree)} \simeq & -\frac{\left(g_Y m_{1D} - \sqrt{2} \lambda_S \mu\right)^2}{m_{SR}^2} - \frac{\left(g_2 m_{2D} + \sqrt{2} \lambda_T  \mu\right)^2}{m_{TR}^2}\nn \\
\delta  \lambda_2 ^{(tree)} \simeq &   -\frac{\left(g_Y m_{1D} + \sqrt{2} \lambda_S \mu\right)^2}{m_{SR}^2} - \frac{\left(g_2 m_{2D} - \sqrt{2} \lambda_T  \mu\right)^2}{m_{TR}^2} \nn\\
\delta \lambda_3^{(tree)}  \simeq &  \, \, \, \,     \frac{g_Y^2 m_{1D}^2 - 2\lambda_S^2 \mu^2}{m_{SR}^2}-  \frac{g_2^2 m_{2D}^2 - 2\lambda_T^2 \mu^2}{m_{TR}^2} \nn  \\
\delta \lambda_4 ^{(tree)} \simeq & \, \, \, \,  \frac{2 g_2^2 m_{2D}^2 - 4 \lambda_T^2 \mu^2}{m_{TR}^2}\,,
\label{deltaLambdaTree}
\end{align}
These induce
 \begin{eqnarray}
\delta V_{4\Phi}^{(tree)}  &=& \delta \lambda_{|0_1,0>}^{(tree)}  |0_1,0\rangle  +   \delta \lambda_{|0_2,0>}^{(tree)}   |0_2,0\rangle  +  \delta \lambda_{|1,0>}^{(tree)}  |1,0\rangle  +   \delta \lambda_{|2,0>}^{(tree)}   |2,0\rangle \, .
\label{deltareparam2HDM} 
\end{eqnarray}
The corrections to the two singlet coefficients 
\begin{align}
\delta \lambda_{|0_1,0>}^{(tree)} \simeq & -2 \lambda_S^2  \frac{\mu^2}{m_{SR}^2} - g_2^2 \frac{ m_{2D}^2}{m_{TR}^2}  \\
\delta  \lambda_{|0_2,0> }^{(tree)}\simeq &   \frac{1}{\sqrt{3}} \left[ g_Y^2 \frac{ m_{1D}^2 }{m_{SR}^2} - 2 g_2^2 \frac{ m_{2D}^2 }{m_{TR}^2}  + 6 \lambda_T^2 \frac{\mu^2}{m_{TR}^2} \right]
\label{deltaLambdaTreeSU2R}
\end{align}
do not contribute to a misalignment. The misalignment arises from the appearance of new terms in the scalar potential:
\begin{align}
\delta \lambda_{|1,0>}^{(tree)} \simeq & 2 g_2 \lambda_T  \frac{m_{2D} \mu}{m_{TR}^2}  -2 g_Y \lambda_S  \frac{m_{1D} \mu}{m_{SR}^2}  \nn \\
\simeq & \sqrt{2} g_2^2  \frac{m_{2D} \mu}{m_{TR}^2}  -\sqrt{2} g_Y^2   \frac{m_{1D} \mu}{m_{SR}^2}  \nn \\
\delta  \lambda_{|2,0> }^{(tree)}\simeq &  \sqrt{ \frac{2}{3}} \left[g_Y^2 \frac{ m_{1D}^2 }{m_{SR}^2} + g_2^2 \frac{ m_{2D}^2 }{m_{TR}^2} \right]
\label{deltaLambdas}
\end{align}
These preserve the subgroup $U(1)_R^{(diag)}$. This is because the scalar potential results from  integrating out the adjoints which have zero $U(1)_R^{(diag)}$ charge. For a numerical estimate, we take  $m_{SR} \simeq m_{TR} \simeq 5$~TeV,   $m_{1D}\simeq m_{1D} \simeq \mu \simeq 500$~GeV,  $g_Y\simeq 0.37$ and $g_2 \simeq 0.64$.  This gives 
\begin{align}
\delta \lambda_{|1,0>}^{(tree)} \simeq & 4 \times 10^{-3},  \qquad
\delta  \lambda_{|2,0> }^{(tree)}\simeq 4.5 \times 10^{-3}&  
\label{deltaLambdaTreeSU2R2}
\end{align}
This shows that this contribution to $Z_6$ can be neglected.

We consider now the misalignment from quantum corrections. Supersymmetry breaking induces mass splitting between scalars and fermionic partners that lead to radiative corrections. 

Loops of the adjoint scalar fields $S$ and $T^a$ do not lead to any contribution as long as their couplings $\lambda_S$ and $\lambda_T$ are given by their $N=2$ values, which is the leading order approximation. This is a consequence of the facts that these scalars are singlets under the $SU(2)_R$ symmetry and at leading order and their interactions with the two Higgs doublets preserve $SU(2)_R$.  The absence of a contribution to $Z_6$ was obtained by explicit calculations of the loop diagrams in Equation (3.5) of \cite{Benakli:2018vqz}. It~was found that when summed up different contributions to $Z_6$ cancel out. This result is now easily understood as a consequence of the $SU(2)_R$ symmetry.

Let's denote by  $D^a$ for the gauge fields $A^a$ and $F_\Sigma^a$ the auxiliary fields for the adjoint scalars $\Sigma^a \in \{ S,T^a \}$ of $U(1)_Y$ and $SU(2)$ respectively. The set:
\begin{align}
( F_\Sigma^a, \quad  \qquad
D^a, \quad  \qquad  {F_\Sigma^a}^*  )
\label{Auxil}
\end{align}
constitutes a triplet of $SU(2)_R$ thus implying the equalities $\lambda_S=  g_Y/{\sqrt{2}}$ and $ \lambda_T = g_2/{\sqrt{2}}$ in  Equation~(\ref{LSTN2}). The violation of these relations by quantum effects translates into breaking of $SU(2)_R$. The correction due to running of the couplings $\lambda_S$ and $ \lambda_T$ leads to a violation of $N=2$ relations (\ref{LSTN2}). This arises first from the radiative corrections from $N=1$ chiral matter. As $\lambda_1$ and $\lambda_2$ are affected in the same way, we have $\delta \lambda_{|1,0>}^{(2\rightarrow 1)} =0$, and using (\ref{lambdaofSU(2)}), we get:
 \begin{eqnarray}
\delta  Z_6^{(2\rightarrow 1)} &=& \frac{\sqrt{6}}{2}  \, \,  s_{2\beta}  \, \,  c_{2\beta}   \, \,  \delta   \lambda_{|2,0>}^{(2\rightarrow 1)} \nn \\
&= &  -  \frac{1}{2}  \, \,\frac{ t_\beta (t_\beta^2-1)}{(1+ t_\beta^2)^2}  \, \,\left[ (2 \lambda_S^2- g_Y^2  )  + ( 2 \lambda_T^2 -g_2^2 ) \right] 
\label{Z6N2-1}
\end{eqnarray}

\begin{figure}[H]
\centering
\includegraphics[width=0.6\textwidth]{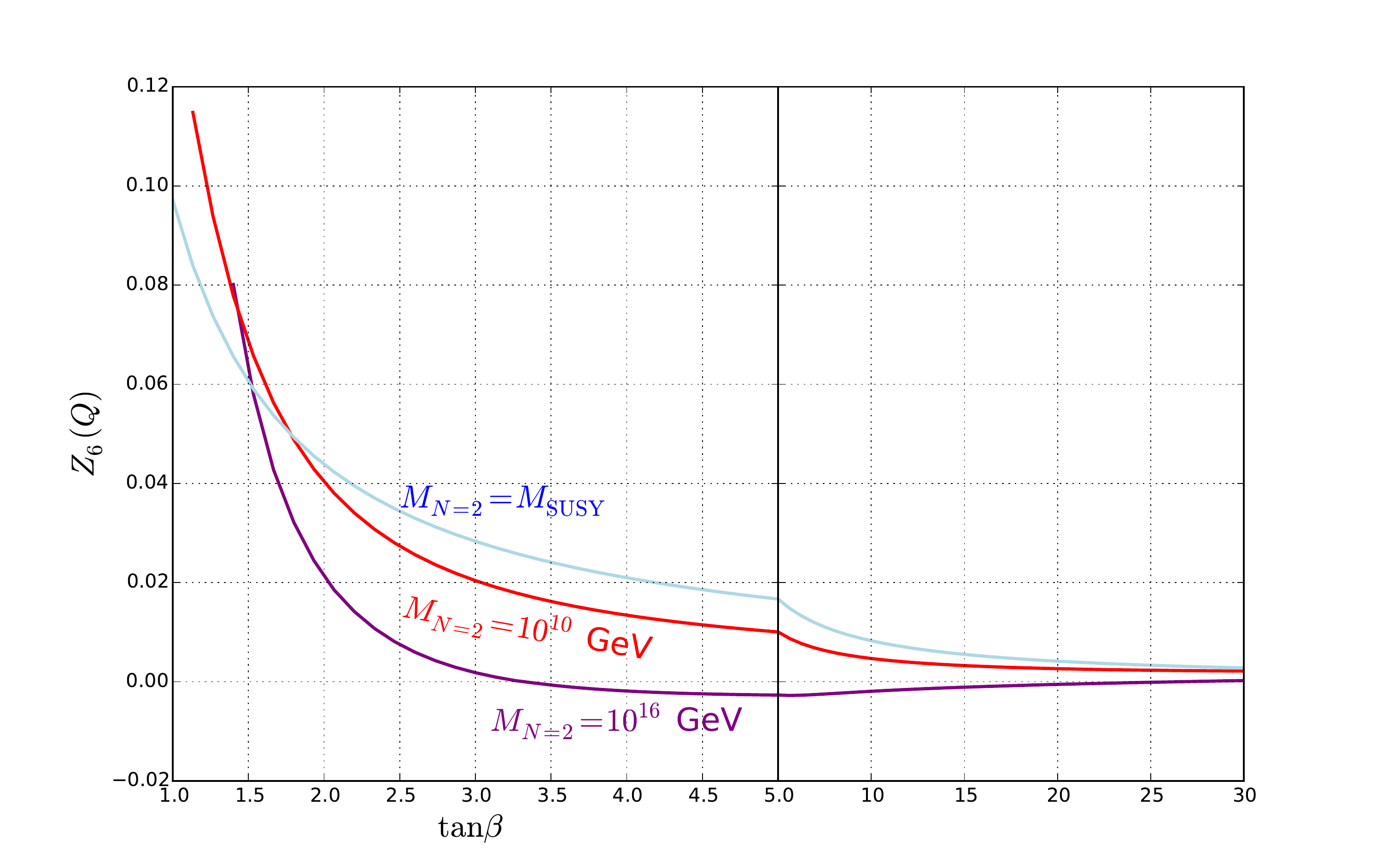}
\caption{$Z_6 (Q)$ at the low energy scale $Q$ against $\tan \beta$ for the $N=2$ scale $M_{N=2} = \MSUSY, 10^{10}$ GeV and $10^{16}$ GeV \cite{Benakli:2018vqz}. }
\label{FIG:Z6}
\end{figure}

In addition to the misalignment from the $N=2 \rightarrow N=1$ described above, there is a contribution from the $N=1 \rightarrow N=0$ mass splitting in chiral superfields. The difference in Yukawa couplings to the two Higgs doublets breaks the $SU(2)_R$ symmetry. For $t_\beta \sim \mathcal{O}(1)$, the biggest contribution is to $\lambda_2$ from stop loops due to their large Yukawa coupling: 
\begin{align}
 \delta  \lambda_2 \sim &  \frac{3y_t^4}{8\pi^2} \log \frac{m_{\tilde{t}}^2}{Q^2} 
\label{toplambda2}
\end{align}
Here $Q$, $y_t$, $m_{\tilde{t}}$ are the renormalisation scale, the top Yukawa coupling and the stop mass, respectively. 
At the end we get:
\begin{align}
Z_6  \approx&\frac{0.12}{t_\beta} - \, \,\frac{ t_\beta ( t_\beta^2-1)}{(1+ t_\beta^2)^2}  \left[  ( 2 \lambda_S^2 - g_Y^2)+ ( 2 \lambda_T^2 - g_2^2) \right].
\end{align}
We find that the misalignment comes from the squark corrections are compensated by the effect of running $\lambda_S, \lambda_T$. The numerical results are shown in Figure \ref{FIG:Z6}, taken from \cite{Benakli:2018vqz}.

\noindent
\section*{Acknowledgments}
{We are grateful to  M.~Goodsell for collaboration on part of the material presented here and P.~Slavich for useful discussions. We acknowledge the support of  the Agence Nationale de Recherche under grant ANR-15-CE31-0002 ``HiggsAutomator''. This work is also supported by the Labex ``Institut Lagrange de Paris'' (ANR-11-IDEX-0004-02,  ANR-10-LABX-63).}

\noindent

\providecommand{\href}[2]{#2}\begingroup\raggedright


\begin{thebibliography}{10}

 %\cite{Gunion:2002zf}
\bibitem{Gunion:2002zf}
Gunion, J.F.,  Haber, H.E.
  {{The CP conserving two Higgs doublet model: The Approach to the decoupling limit}}?
   {\color{blue} {\em Phys.\ Rev.\ D } {\bf2003}, {\color{blue} \emph {67}, 075019, }}%doi:10.1103/PhysRevD.67.075019

  %%CITATION = }}%doi:10.1103/PhysRevD.67.075019, %%



\bibitem{Antoniadis:2006uj}
Antoniadis, I.,  Benakli, K.,  Delgado, A.,  Quiros, M.  {  {A New gauge
  mediation theory}}. {\color{blue} {\em Adv. Stud. Theor. Phys. } {\bf2008}, {\color{blue} \emph {2},  645--672.}}
%%CITATION = HEP-PH/0610265, %%.


\bibitem{Ellis:2016gxa}
Ellis, J.,  Quevillon, J.,  Sanz, V. { {Doubling Up on Supersymmetry in the
  Higgs Sector}}. {\color{blue} \emph {J. High Energy Phys.}  {\bf2016}, {\color{blue} \emph {10},   086, }}%doi:10.1007/JHEP10(2016)086.
%%CITATION = ARXIV:1607.05541, %%.

\bibitem{Benakli:2018vqz}
Benakli, K.,  Goodsell, M.D.,  Williamson, S.L. { {Higgs alignment from
  extended supersymmetry}}. {\color{blue} \emph {Eur. Phys. J. C}  {\bf2018}, {\color{blue} \emph {78},   658, }}%doi:10.1140/epjc/s10052-018-6125-1.
%%CITATION = ARXIV:1801.08849, %%.

%\cite{Benakli:2018vjk}
\bibitem{Benakli:2018vjk}
 Benakli, K.,  Chen, Y.,  Lafforgue-Marmet, G.
 { {R-symmetry for Higgs alignment without decoupling}}. {\color{blue} \emph {arXiv}  \textbf{2018},    arXiv:1811.08435.}
% \href{http://arxiv.org/abs/1811.08435}{{\tt [hep-ph]}}.
  %%CITATION = ARXIV:1811.08435, %%
   
\bibitem{Belanger:2009wf}
Belanger, G.,  Benakli, K.,  Goodsell, M.,  Moura, C.,  Pukhov, A. { {Dark
  Matter with Dirac and Majorana Gaugino Masses}}. {\color{blue} \emph {J. Cosmol. Astropart. Phys. }    {\bf2009}, }%doi:10.1088/1475-7516/2009/08/027.
%%CITATION = ARXIV:0905.1043, %%.

\bibitem{Gunion:1989we}
Gunion, J.F.,  Haber, H.E.,  Kane, G.L.,  Dawson, S. { {The Higgs Hunter's
  Guide}}.
{\color{blue} \emph {Front. Phys.} {\bf 2000},  {\em 80},   1--404.}
%%CITATION = FRPHA,80,1, %%.

\bibitem{Branco:2011iw}
Branco, G.C.,  Ferreira, P.M.,  Lavoura, L.,  Rebelo, M.N.,  Sher, M.,  Silva, J.P. { {Theory and phenomenology of two-Higgs-doublet models}}. {\color{blue} \emph {Phys. Rept. }     {\bf 2012}, {\color{blue} {\em 516},   }}%doi:10.1016/j.physrep.2012.02.002.
%%CITATION = ARXIV:1106.0034, %%.

\bibitem{Djouadi:2005gj}
Djouadi, A. { {The Anatomy of electro-weak symmetry breaking. II. The Higgs
  bosons in the minimal supersymmetric model}}.  {\color{blue} \emph {Phys. Rept. }     {\bf 2008}, {\color{blue} {\em 459}, }}%doi:10.1016/j.physrep.2007.10.005.
%%CITATION = HEP-PH/0503173, %%.

\bibitem{Bernon:2015qea}
Bernon, J.,  Gunion, J.F.,  Haber, H.E.,  Jiang, Y.,  Kraml, S. {
  {Scrutinizing the alignment limit in two-Higgs-doublet models: m$_h$~=~125
  GeV}}.  {\color{blue} \emph {Phys. Rev. D}     {\bf 2015}, {\color{blue} {\em 92}, 075004, }}%doi:10.1103/PhysRevD.92.075004.
%%CITATION = ARXIV:1507.00933, %%.

\bibitem{Bernon:2015wef}
Bernon, J.,  Gunion, J.F.,  Haber, H.E.,  Jiang, Y.,  Kraml, S. {
  {Scrutinizing the alignment limit in two-Higgs-doublet models. II.
  m$_H$ = 125 GeV}}.    {\color{blue} \emph {Phys. Rev. D}     {\bf 2016}, {\color{blue} {\em 93},  035027, 
  }}%doi:10.1103/PhysRevD.93.035027.
%%CITATION = ARXIV:1511.03682, %%.

\bibitem{Carena:2013ooa}
Carena, M.,  Low, I.,  Shah, N.R.,  Wagner, C.E.M. { {Impersonating the
  Standard Model Higgs Boson: Alignment without Decoupling}}.  {\color{blue} \emph {J. High Energy Phys.}  {\bf2014}, {\color{blue} \emph {4},   
   015, }}%doi:10.1007/JHEP04(2014)015.
%%CITATION = ARXIV:1310.2248, %%.

\bibitem{Carena:2015moc}
Carena, M.,  Haber, H.E.,  Low, I.,  Shah, N.R.,  Wagner, C.E.M. {
  {Alignment limit of the NMSSM Higgs sector}}.  {\color{blue} \emph {Phys. Rev. D}     {\bf 2016}, {\color{blue} {\em 93},   035013, 
 }}%doi:10.1103/PhysRevD.93.035013.
%%CITATION = ARXIV:1510.09137, %%.

\bibitem{Haber:2017erd}
Haber, H.E.,  Heinemeyer, S.,  Stefaniak, T. { {The Impact of Two-Loop
  Effects on the Scenario of MSSM Higgs Alignment without Decoupling}}.  {\color{blue} \emph {Eur. Phys. J. C}  {\bf2017}, {\color{blue} \emph {77},  742,  }}%doi:10.1140/epjc/s10052-017-5243-5.
%%CITATION = ARXIV:1708.04416, %%.

 
\bibitem{Davidson:2005cw}
Davidson, S.,  Haber, H.E. {{Basis-independent methods for the
  two-Higgs-doublet model}}. {\color{blue} \emph {Phys. Rev. D}     {\bf 2005}, {\color{blue} {\em 72},    035004, }}%doi:10.1103/PhysRevD.72.099902.
 
%%CITATION = HEP-PH/0504050, %%.

\bibitem{Ivanov:2005hg}
Ivanov, I.P. { {Two-Higgs-doublet model from the group-theoretic
  perspective}}.  {\color{blue} \emph {Phys.  Lett. B}   {\bf 2006}, {\color{blue} {\em 632},  }}%doi:10.1016/j.physletb.2005.10.015.
%%CITATION = HEP-PH/0507132, %%.

\bibitem{Ferreira:2009wh}
Ferreira, P.M.,  Haber, H.E.,  Silva, J.P. { {Generalized CP symmetries
  and special regions of parameter space in the two-Higgs-doublet model}}.  {\color{blue} \emph {Phys. Rev. D}     {\bf 2009}, {\color{blue} {\em 79}, 116004, }}%doi:10.1103/PhysRevD.79.116004. 
 
%%CITATION = ARXIV:0902.1537, %%.

\bibitem{Dev:2014yca}
Dev, P.S.B.,  Pilaftsis, A. { {Maximally Symmetric Two Higgs Doublet
  Model with Natural Standard Model Alignment}}.  {\color{blue} \emph {J. High Energy Phys.}  {\bf2014}, {\color{blue} \emph {12}, 024, }}%doi:10.1007/JHEP11(2015)147.
  
%%CITATION = ARXIV:1408.3405, %%.

\bibitem{Lane:2018ycs}
Lane, K.,  Shepherd, W. { {Natural Stabilization of the Higgs Boson's Mass
  and Alignment}}.  {\color{blue} \emph {arXiv}  \textbf{2018},  arXiv:1808.07927.}
%\href{http://arxiv.org/abs/1808.07927}{{\tt  [hep-ph]}}.
%%CITATION = ARXIV:1808.07927, %%.


\bibitem{Ferreira:2010bm}
Ferreira, P.M.,  Silva, J.P. { {A Two-Higgs Doublet Model With Remarkable
  CP Properties}}.   {\color{blue} \emph {Eur. Phys. J. C}  {\bf2010}, {\color{blue} \emph {69},    45--52, }}%doi:10.1140/epjc/s10052-010-1384-5.
%%CITATION = ARXIV:1001.0574, %%.

\bibitem{Fayet:1975yi}
Fayet, P. { {Fermi-Bose Hypersymmetry}}. {\color{blue} \emph {Nucl. Phys. B }     {\bf 1976}, {\color{blue} \emph {113},   135,  }}%doi:10.1016/0550-3213(76)90458-2.
%%CITATION = NUPHA,B113,135, %%.

\bibitem{delAguila:1984qs}
del Aguila, F.,  Dugan, M.,  Grinstein, B.,  Hall, L.J.,  Ross, G.G.,  West, P.C.
  { {Low-energy Models With Two Supersymmetries}}.  {\color{blue} \emph {Nucl. Phys. B }     {\bf 1985}, {\color{blue} \emph {250},    225--251, }}%doi:10.1016/0550-3213(85)90480-8.
%%CITATION = NUPHA,B250,225, %%.

\bibitem{Antoniadis:2005em}
Antoniadis, I.,  Delgado, A.,  Benakli, K.,  Quiros, M.,  Tuckmantel, M. {
  {Splitting extended supersymmetry}}.    {\color{blue} \emph {Phys. Rev. B}     {\bf 2006}, {\color{blue} {\em 634}, 302--306, }}%doi:10.1016/j.physletb.2006.01.010.
%%CITATION = HEP-PH/0507192, %%.

\bibitem{Antoniadis:2006eb}
Antoniadis, I.,  Benakli, K.,  Delgado, A.,  Quiros, M.,  Tuckmantel, M. {
  {Split extended supersymmetry from intersecting branes}}.    {\color{blue} \emph {Nucl. Phys. B }     {\bf 2006}, {\color{blue} \emph {744},   156--179, }}%doi:10.1016/j.nuclphysb.2006.03.012.
%%CITATION = HEP-TH/0601003, %%.

\bibitem{Allanach:2006fy}
Benakli, K.,  Moura, C. {{Les Houches physics at TeV colliders 2005 beyond
  the standard model working group: Summary report}}.    {\color{blue} \emph {arXiv}  \textbf{2006}, arXiv:hep-ph/0602198.}

%%CITATION = HEP-PH/0602198, %%.

\bibitem{Fayet:1978qc}
Fayet, P. {{Massive gluinos}}. {\color{blue} \emph {Phys. Lett. B}  {\bf 1978}, {\color{blue} \emph {78},   417--420.}}
%%CITATION = PHLTA,78B,417, %%.

\bibitem{Polchinski:1982an}
Polchinski, J.,  Susskind, L. {{Breaking of Supersymmetry at
  Intermediate-Energy}}.   {\color{blue} \emph {Phys. Rev. D}     {\bf 1982}, {\color{blue} {\em 26},   3661, }}%doi:10.1103/PhysRevD.26.3661.
%%CITATION = PHRVA,D26,3661, %%.

\bibitem{Hall:1990hq}
Hall, L.J.,  Randall, L. { {U(1)-R symmetric supersymmetry}}. {\color{blue} \emph {Nucl. Phys. B }     {\bf 1991}, {\color{blue} \emph {352},   289--308, }}%doi:10.1016/0550-3213(91)90444-3.
%%CITATION = NUPHA,B352,289, %%.

\bibitem{Fox:2002bu}
Fox, P.J.,  Nelson, A.E.,  Weiner, N. { {Dirac gaugino masses and
  supersoft supersymmetry breaking}}.  {\color{blue} \emph {J. High Energy Phys.} {\bf 2002}, {\color{blue} \emph {8}, 035, }}%doi:10.1088/1126-6708/2002/08/035. 
  
%%CITATION = HEP-PH/0206096, %%.

\bibitem{Benakli:2008pg}
Benakli, K.,  Goodsell, M.D. { {Dirac Gauginos in General Gauge
  Mediation}}.   {\color{blue} \emph {Nucl. Phys. B }     {\bf 2009}, {\color{blue} \emph {816}, 185--203, }}%doi:10.1016/j.nuclphysb.2009.03.002.
%%CITATION = ARXIV:0811.4409, %%.

\bibitem{Benakli:2009mk}
Benakli, K.,  Goodsell, M.D. { {Dirac Gauginos and Kinetic Mixing}}.       {\color{blue} \emph {Nucl. Phys. B }     {\bf 2010}, {\color{blue} \emph {830}, 315--329, }}%doi:10.1016/j.nuclphysb.2010.01.003. 
 

\bibitem{Amigo:2008rc}
Amigo, S.D.L.,  Blechman, A.E.,  Fox, P.J.,  Poppitz, E. { {R-symmetric
  gauge mediation}}.  {\color{blue} \emph {J. High Energy Phys.}    {\bf 2009}, {\color{blue} \emph {1},   018, }}%doi:10.1088/1126-6708/2009/01/018. 
  
%%CITATION = ARXIV:0809.1112, %%.

\bibitem{Benakli:2010gi}
Benakli, K.,  Goodsell, M.D. { {Dirac Gauginos, Gauge Mediation and
  Unification}}.  {\color{blue} \emph {Nucl. Phys. B }     {\bf 2010}, {\color{blue} \emph {840},  1--28, }}%doi:10.1016/j.nuclphysb.2010.06.018.
%%CITATION = ARXIV:1003.4957, %%.

\bibitem{Choi:2010gc} Choi, S.Y.,  Choudhury, D.,  Freitas, A.,  Kalinowski, J.,  Kim, J.M.,  Zerwas, P.M. { {Dirac Neutralinos and Electroweak Scalar Bosons of N = 1/N = 2
  Hybrid Supersymmetry at Colliders}}.  {\color{blue} \emph {J. High Energy Phys.} {\bf 2010}, {\color{blue} \emph {8},   025, }}%doi:10.1007/JHEP08(2010)025.
%%CITATION = ARXIV:1005.0818, %%.

\bibitem{Benakli:2011vb}
Benakli, K. {{Dirac Gauginos: A User Manual}}.    {\color{blue} \emph {  Fortsch. Phys. }   {\bf 2011}, {\color{blue} \emph {59},    1079--1082, }}%doi:10.1002/prop.201100071.
%%CITATION = ARXIV:1106.1649, %%.

\bibitem{Benakli:2011kz}
Benakli, K.,  Goodsell, M.D.,  Maier, A.-K. { {Generating mu and Bmu in
  models with Dirac Gauginos}}.    {\color{blue} \emph {Nucl. Phys. B }     {\bf 2011}, {\color{blue} \emph {851},  445--461, }}%doi:10.1016/j.nuclphysb.2011.06.001. 
%%CITATION = ARXIV:1104.2695, %%.

\bibitem{Itoyama:2011zi}Itoyama, H.,  Maru, N. { {D-term Dynamical Supersymmetry Breaking
  Generating Split N=2 Gaugino Masses of Mixed Majorana-Dirac Type}}.  {\color{blue} \emph {Int. J. Mod. Phys. A}  {\bf 2012}, {\color{blue} \emph {27},  1250159, }}%doi:10.1142/S0217751X1250159X.
  

\bibitem{Benakli:2012cy}
Benakli, K.,  Goodsell, M.D.,  Staub, F. { {Dirac Gauginos and the 125 GeV
  Higgs}}.  {\color{blue} \emph {J. High Energy Phys.}   {\bf 2013}, {\color{blue} \emph {6},   073, }}%doi:10.1007/JHEP06(2013)073.
%%CITATION = ARXIV:1211.0552, %%.

\bibitem{Benakli:2014cia}
Benakli, K.,  Goodsell, M.,  Staub, F.,  Porod, W. { {Constrained minimal
  Dirac gaugino supersymmetric standard model}}.   {\color{blue} \emph {    Phys. Rev.  D}  {\bf 2014}, {\color{blue} \emph {90},  045017,  }}%doi:10.1103/PhysRevD.90.045017.
%%CITATION = ARXIV:1403.5122, %%.

\bibitem{Martin:2015eca}
Martin, S.P. { {Nonstandard supersymmetry breaking and Dirac gaugino masses
  without supersoftness}}.      {\color{blue} \emph {Phys. Rev.  D}  {\bf 2015}, {\color{blue} \emph {92},   035004, }}%doi:10.1103/PhysRevD.92.035004.
%%CITATION = ARXIV:1506.02105, %%.

\bibitem{Braathen:2016mmb}Braathen, J.,  Goodsell, M.D.,  Slavich, P.     {Leading two-loop corrections
  to the Higgs boson masses in SUSY models with Dirac gauginos}.  {\color{blue} \emph {J. High Energy Phys.}    {\bf 2016}, {\color{blue} \emph {9},      045,  }}%doi:10.1007/JHEP09(2016)045.
%%CITATION = ARXIV:1606.09213, %%.

%\cite{Unwin:2012fj}
\bibitem{Unwin:2012fj}
  Unwin, J.
  R-symmetric High Scale Supersymmetry.        {\color{blue} \emph {Phys. Rev.  D}  {\bf 2012}, {\color{blue} \emph {86}, 095002,  }}%doi:10.1103/PhysRevD.86.095002.
  
  %%CITATION = }}%doi:10.1103/PhysRevD.86.095002, %%
  
  %\cite{Chakraborty:2018izc}
\bibitem{Chakraborty:2018izc}
Chakraborty, S.,  Martin, A.,  Roy, T.S.
  Charting generalized supersoft supersymmetry.   {\color{blue} \emph {J. High Energy Phys.}    {\bf 2018}, {\color{blue} \emph {1805}, 176,  }}%doi:10.1007/JHEP05(2018)176. 

  %%CITATION = }}%doi:10.1007/JHEP05(2018)176, %%
  
  \bibitem{Csaki:2013fla}Csaki, C.,  Goodman, J.,  Pavesi, R.,  Shirman, Y. { {The $m_D-b_M$ problem of
  Dirac gauginos and its solutions}}.             {\color{blue} \emph {Phys. Rev.  D}  {\bf 2014}, {\color{blue} \emph {89}, 055005, }}%doi:10.1103/PhysRevD.89.055005. 
 
%%CITATION = ARXIV:1310.4504, %%.


\bibitem{Benakli:2016ybe}{Benakli, K}.,  Darm\'e, L.,  Goodsell, M.D.,  Harz, J. { {The Di-Photon Excess
  in a Perturbative SUSY Model}}.    {\color{blue} \emph {Nucl. Phys. B }     {\bf 2016}, {\color{blue} \emph {911}, 127--162, }}%doi:10.1016/j.nuclphysb.2016.07.027.
%%CITATION = ARXIV:1605.05313, %%.

\bibitem{Nelson:2015cea}{Nelson, A.E}.,  Roy, T.S. { {New Supersoft Supersymmetry Breaking
  Operators and a Solution to the $\mu$ Problem}}.      {\color{blue} \emph { Phys. Rev. Lett. }         {\bf 2015}, {\color{blue} \emph {114}, 201802, 
  }}%doi:10.1103/PhysRevLett.114.201802.
%%CITATION = ARXIV:1501.03251, %%.

\bibitem{Alves:2015kia}{Alves, D.S.M., } Galloway, J.,  McCullough, M.,  Weiner, N. { {Goldstone
  Gauginos}}.   {\color{blue} \emph { Phys. Rev. Lett. }         {\bf 2015}, {\color{blue} \emph {115}, 161801, }}%doi:10.1103/PhysRevLett.115.161801.
%%CITATION = ARXIV:1502.03819, %%.

\bibitem{Alves:2015bba}{Alves, D.S.M}.,  Galloway, J.,  McCullough, M.,  Weiner, N.   { {Models of
  Goldstone Gauginos}}.                {\color{blue} \emph {Phys. Rev.  D}  {\bf 2016}, {\color{blue} \emph {93}, 075021, }}%doi:10.1103/PhysRevD.93.075021.
%%CITATION = ARXIV:1502.05055, %%.

\bibitem{Haber:1993an}Haber, H.E.,  Hempfling, R. { {The Renormalization group improved Higgs
  sector of the minimal supersymmetric model}}.    {\color{blue} \emph {Phys. Rev.  D}  {\bf 1993}, {\color{blue} \emph {48},     4280--4309, }}%doi:10.1103/PhysRevD.48.4280.
%%CITATION = HEP-PH/9307201, %%.

\end{thebibliography}
\end{document}